\documentclass[12pt,twoside]{article}   
\usepackage[super,sort,comma]{natbib}
\usepackage{amssymb}

\usepackage{fancyhdr}		

\usepackage{showkeys}		

\newcommand{\captionv}[3]{\begin{center}\parbox{#1cm}{\caption[#2]{{\sf #3}}}
        \end{center}}

\usepackage[section]{placeins}   %

\usepackage{graphicx}

\makeatletter \renewcommand\@biblabel[1]{$^{#1}$} \makeatother
\newcommand{\cen}[1]{\begin{center} #1 \end{center}}

\usepackage[mathlines]{lineno}

\usepackage{hyperref}
\hypersetup{ colorlinks,
    citecolor=blue,
    filecolor=blue,
    linkcolor=blue,
    urlcolor=blue
}

\usepackage{xcolor}

\definecolor{gray}{rgb}{0.6,0.6,0.6}
\definecolor{red}{rgb}{0.85,0,0}
\definecolor{green}{rgb}{0,0.85,0}
\definecolor{blue}{rgb}{0,0,0.85}
\definecolor{beige}{rgb}{0.92,0.87,0.78}
\usepackage[all]{hypcap}    

\begin{document}

\cen{\sf {\Large {\bfseries Real-time interactive magnetic resonance (MR) temperature imaging in both aqueous and adipose tissues using cascaded deep neural networks for MR-guided focused ultrasound surgery (MRgFUS) } \\  
\vspace*{10mm}
Jong-Min Kim} \\
Department of Electronics and Information Engineering, Korea University, Sejong 30019, Republic of Korea \\
Korea Artificial Organ Center, Korea University, Seoul 02481, Republic of Korea \\
ICT Convergence Technology for Health and Safety, Korea University, Sejong 30019, Republic of Korea \\
{\Large
\vspace*{10mm}
You-Jin Jeong} \\ 
Department of Electronics and Information Engineering, Korea University, Sejong 30019, Republic of Korea \\
Korea Artificial Organ Center, Korea University, Seoul 02481, Republic of Korea \\
ICT Convergence Technology for Health and Safety, Korea University, Sejong 30019, Republic of Korea \\
{\Large
\vspace*{10mm}
Han-Jae Chung} \\
Department of Electronics and Information Engineering, Korea University, Sejong 30019, Republic of Korea \\
Korea Artificial Organ Center, Korea University, Seoul 02481, Republic of Korea \\
ICT Convergence Technology for Health and Safety, Korea University, Sejong 30019, Republic of Korea \\
{\Large
\vspace*{10mm}
Chulhyun Lee} \\ 
Bioimaging Research Team, Korea Basic Science Institute, Cheongju 43133, Republic of Korea \\
{\Large
\vspace*{10mm}
Chang-Hyun Oh$^{a)}$} \\ 
Department of Electronics and Information Engineering, Korea University, Sejong 30019, Republic of Korea \\
Korea Artificial Organ Center, Korea University, Seoul 02481, Republic of Korea \\
ICT Convergence Technology for Health and Safety, Korea University, Sejong 30019, Republic of Korea \\
\vspace{5mm} 
Version typeset 19-July-2019}

\pagenumbering{roman}
\setcounter{page}{1}
\pagestyle{plain}
$^{a)}$Author to whom correspondence should be addressed. email: ohch@korea.ac.kr;
R and D center, Room number 337, Nokji-Campus, Korea University, Anam-ro 145, Seongbuk-gu, Seoul, 02841, Korea; Telephone: +82-2-3290-3984; Fax: +82-2-3290-4294\\

\begin{abstract}

\noindent {\bf Purpose:} Real-time interactive magnetic resonance (MR) temperature mapping of aqueous and adipose tissue is a challenging technique for guiding thermal therapy. To acquire the real-time interactive temperature map, the problems of long acquisition and processing time must be addressed. However, there are in turn three major considerations when it comes to reducing the acquisition and processing times. First, the temperature of adipose and aqueous tissue must be calculated in a different manner; temperature images for adipose and aqueous tissue are calculated by T1 and proton resonance frequency (PRF)-based MR thermometry, respectively. Second, the acquisition time itself should be reduced. Third, the time for processing the low-resolution (LR) MR images to high-resolution (HR) MR images and calculating the T1 relaxation map should be reduced. To overcome these major challenges, this paper proposes a cascaded convolutional neural network (CNN) framework and multi-echo gradient echo (meGRE) with a single reference variable flip angle (srVFA).\\
{\bf Methods:} To optimize the echo times for each method, MR images are acquired using a meGRE sequence; meGRE images with two flip angles (FAs) and meGRE images with a single FA are acquired during the pretreatment and treatment stages, respectively. These images are then processed and reconstructed by a cascaded CNN, which consists of two CNNs. The first CNN (called DeepACCnet) performs HR complex MR image reconstruction from the LR MR image acquired during the treatment stage, which is improved by the HR magnitude MR image acquired during the pretreatment stage. The second CNN (called DeepPROCnet) copes with T1 mapping. The DeepACCnet-generated HR magnitude MR image with higher FA and the HR magnitude MR image with lower FA acquired during the pretreatment stage are used as the input, and T1 reconstructed by srVFA is used as the label. The DeepACCnet-generated HR phase MR image with higher FA and the pretreatment-HR phase MR image with higher FA are used to calculate the temperature map, which is based on PRF-based MR thermometry.\\
{\bf Results:} Measurements of temperature and T1 changes obtained by meGRE combined with srVFA and cascaded CNNs were achieved in an agarose gel phantom, ex vivo porcine muscle, and ex vivo porcine muscle with fat layers (heating tests), and in vivo human prostate and brain (non-heating tests). In the heating test, the maximum differences between fiber-optic sensor and samples are less than 1 $^{\circ}$C. In all cases, temperature mapping using the cascaded CNN achieved the best results in all cases. The acquisition and processing times for the proposed method are 0.8 s and 32 ms, respectively. \\
{\bf Conclusions:} Real-time interactive HR MR temperature mapping for simultaneously measuring aqueous and adipose tissue is feasible by combining a cascaded CNN with meGRE and srVFA.  

\end{abstract}

\newpage     

\tableofcontents

\newpage

\setlength{\baselineskip}{0.7cm}      

\pagenumbering{arabic}
\setcounter{page}{1}
\pagestyle{fancy}

\section{Introduction}

Various imaging modalities, including computed tomography (CT), ultrasound (US), and magnetic resonance imaging (MRI), have been used for image-guided surgery. Among them, CT and US do not allow quantitative temperature imaging. However, MRI allows measurement of the temperature image and calculation of the thermal dose in a variety of tissue types. This is the most important advantage of magnetic resonance (MR)-guided surgery. However, MRI has many limitations in real-time treatment guidance owing to its relatively long acquisition time. There have been several efforts made to solve the issue of long image acquisition time. Because long image acquisition time is not only a problem in thermal treatment guidance but also in clinical magnetic resonance (MR) applications, an attempt to solve the problem by using existing methods to improve the image acquisition time of temperature images was documented\cite{HanYH2011_JMRI}. The basic method used to overcome the long acquisition time is to acquire part of the k-space and then reconstruct the full k-space from the part obtained\cite{Pruessmann1999_MRM}\cite{Griswold2002_MRM}\cite{Uecker2014_MRM}. Recently, compressed sensing (CS) MRI has been widely investigated \cite{Lustig2007_MRM}. After the sparse acquisition of MR images, the CS algorithm allows for the reconstruction of a unique solution and has successfully achieved higher speeds in many applications\cite{Jaspan2015_BrJRadiol}. However, because the nonlinear optimization process requires iterative computing, most CS algorithms have been developed for a two-dimensional (2D) image or dynamic process (e.g., cardiac function and dynamic contrast-enhanced imaging), in which a long latency time does not matter\cite{Kerr1997_MRM}\cite{Nayak2019_MRM}. Therefore, there are limitations to the direct use of CS algorithms in reconstructing temperature images for thermal therapy guidance that requires real-time feedback. 

Although a model-based approach for proton resonance frequency (PRF)-based MR thermometry has been developed to address this, the model-based approach is not applicable to adipose tissue because it achieves high accuracy only when the fat content is low\cite{Poorman2019_MRM}. Therefore, several alternative methods to monitor the temperature changes in adipose tissue, including the T1 and T2 relaxation times, have been proposed\cite{Parker1984_IEEETBE}\cite{Gandhi1984_ISMRM}. Although the T1 change with changing temperature is dependent on the tissue type, it can be used in adipose tissue because the T1 changes linearly in fat from 10 $^{\circ}$C to 70 $^{\circ}$C\cite{Kuroda2011_MRMS}. In particular, simultaneous T1 and PRF-shift measurement methods have been proposed to measure the temperature of aqueous and adipose tissues using the variable flip angle (VFA). In most methods, the PRF-shift is calculated for every acquisition, and the T1 change is calculated by combining sequential images with the VFA method\cite{Hey2012_MRM}\cite{Todd2013_MRM}. Recently, a single reference VFA (srVFA) method was proposed to enhance the scan time and scanning efficiency of these methods\cite{Svedin2019_MRM}. However, because the processing for T1 extraction by srVFA or VFA takes several hours, real-time interactive temperature measurement in adipose tissue remains difficult. 
 
Recently, deep learning, particularly using convolutional neural networks (CNNs), has been applied to reconstruct high-resolution (HR) MR images from low-resolution (LR) MR images or to extract the MR parameters \cite{Kim2018_MedPhys}\cite{Kwon2017_MedPhys}\cite{Akakaya2019_MRM}\cite{Eo2018_MRM} \cite{Hammernik2018_MRM}\cite{Cai2018_MRM}\cite{Liu2019_MRM}\cite{Gibbons2019_MRM}\cite{Domsch2018_MRM}\cite{Cho2019_MRM}. Not only did these CNNs show impressive performance, but processing time was also close to real time\cite{Kwon2017_MedPhys}\cite{Cho2019_MRM}\cite{Yoon2018_NI}, making real-time interactive MR temperature imaging more practical. Considering the current outcomes of the use of CNNs for image reconstruction and MR parameter extraction, this approach may be applicable for MR temperature imaging. In addition, the incorporation of anatomical information from HR MR images can be used to improve CNN MR image reconstruction from LR MR images\cite{Kim2018_MedPhys}. This method is useful for multi-contrast MR image reconstruction, which acquires multiple MRI sequences within a clinical setup, but is limited to single-contrast MR image reconstruction. However, when performing MR-guided thermal therapy, there is a pretreatment stage for obtaining patient position information, calibration, and so on, and there is a treatment stage where this prior information is used to treat a patient. As previously mentioned, the srVFA method is used to obtain images with two flip angles (FAs) during the pretreatment stage and this prior information is used to simultaneously observe the T1 and PRF-shift during the treatment stage. Moreover, MR image reconstruction from an LR MR image acquired during the treatment stage may be applicable to MR image reconstruction by taking advantage of anatomical information from the HR MR image acquired during the pretreatment stage. However, MR temperature mapping with deep learning has not been explored thoroughly to date.

In this study, we describe a real-time interactive MR temperature imaging method for both aqueous and adipose tissues using cascaded CNNs, which consists of one CNN to provide HR complex MR image reconstruction, and one CNN for T1 mapping. We will refer to these neural networks as DeepACCnet (CNN for HR complex MR image reconstruction) and DeepPROCnet (CNN for T1 mapping) hereafter. Every MR-guided focused ultrasound (FUS) requires a short latency period because temperature information obtained from an MRI must be reflected immediately in the FUS treatment (i.e., real-time interactive temperature imaging), which will enable real-time image acquisition and processing through the two CNNs. This novel temperature imaging method is evaluated in MRgFUS experiments using an agarose gel phantom, ex vivo porcine muscle samples, and in vivo human volunteers.

\section{Materials and methods}

In this section, we describe the formulation of the proposed method and the overall data processing in detail.

\subsection{Treatment monitoring strategy: MR pulse sequence and sampling strategy}

\begin{figure}[ht]
   \begin{center}
   \includegraphics[width=14cm]{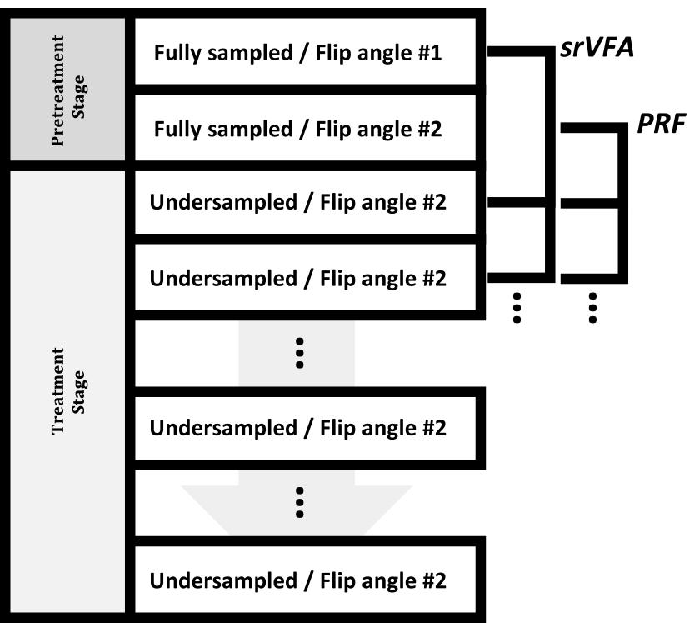}

   %
   %
   \captionv{12}{Short title - can be blank}{The proposed data acquisition scheme for real-time interactive temperature imaging in both adipose and aqueous tissues. During the pretreatment stage, the fully sampled k-space with two flip angles (FAs) is acquired and the undersampled k-space with a single FA is obtained to quickly obtain MR images during the treatment stage. The T1 map by single reference variable flip angle (srVFA) is calculated by using MR images at the lower FA acquired during the pretreatment stage and at the higher FA acquired during the treatment stage. The temperature map by proton resonance frequency-based MR thermometry is calculated by using MR images at the higher FA acquired during the pretreatment stage and at the higher FA acquired during the treatment stage. srVFA: single reference variable flip angel, PRF: proton resonance frequency.
   \label{Fig1} 
    }  
    \end{center}
\end{figure}

In general, T1 is calculated from the magnitude MR images obtained by the spoiled gradient echoes with two FAs\cite{Fram1987_MRI}, and the PRF-shift is obtained from the phase change in the spoiled gradient echo\cite{Ishihara1995_MRM}. To obtain the T1 changes, these methods are not practical in real-time temperature mapping because they must obtain two spoiled gradient echoes. To overcome this, srVFA was proposed\cite{Svedin2019_MRM}. Briefly, T1 mapping by srVFA involves acquiring an MR image at the lower FA during the pretreatment stage and then acquiring MR images at the higher FA during the treatment stage. However, because the MR image acquired during the pretreatment stage does not change, the apparent T1 value determined by srVFA includes a systematic error. If we can determine the true T1 value during the pretreatment stage, the true T1 acquired during the treatment stage can be calculated by applying a simple correction\cite{Svedin2019_MRM}. Another consideration is the echo time (TE). Although the PRF-shift achieves the optimal signal-to-noise ratio (SNR) when T2* and TE are equal, the VFA method achieves optimal SNR when TE is minimized. Therefore, meGRE with srVFA was used to optimize the TE for each method. 
Additionally, an undersampled MR image with a single FA was obtained to accelerate the images acquired during the treatment stage. On the other hand, in the pretreatment stage, fully sampled images with two FAs were acquired (Fig. \ref{Fig1}). 

The complex MR images acquired during the pretreatment stage, $t_{0}$, in $TE_{l}$ are $S_{F
}(FA_{1}, t_{0}; TE_{l})$ $\in \mathbb{C} ^{N_{x} \times N_{y}}$ and $S_{F}(FA_{2}, t_{0};TE_{l})$ $\in \mathbb{C} ^{N_{x} \times N_{y} }$. The complex MR image acquired during the treatment stage at $t_{i}$ in $TE_{l}$ is $S_{U}(FA_{2}, t_{i};TE_{l})$ $\in \mathbb{C} ^{N_{x} \times N_{y} }$. $N_{x}$ and $N_{y}$ denote the number of phase encodings and the number of frequency encodings, respectively. The undersampled k-space acquired during the treatment stage is filled with zeros. During the pretreatment stage, the B1 map, $B1(t_{0})$ $\in \mathbb{C} ^{N_{x} \times N_{y}}$, is optionally acquired for correcting inaccurate FA distributions. 

\subsection{Cascaded convolutional neural networks}

\begin{figure}[ht]
   \begin{center}
   \includegraphics[width=14cm]{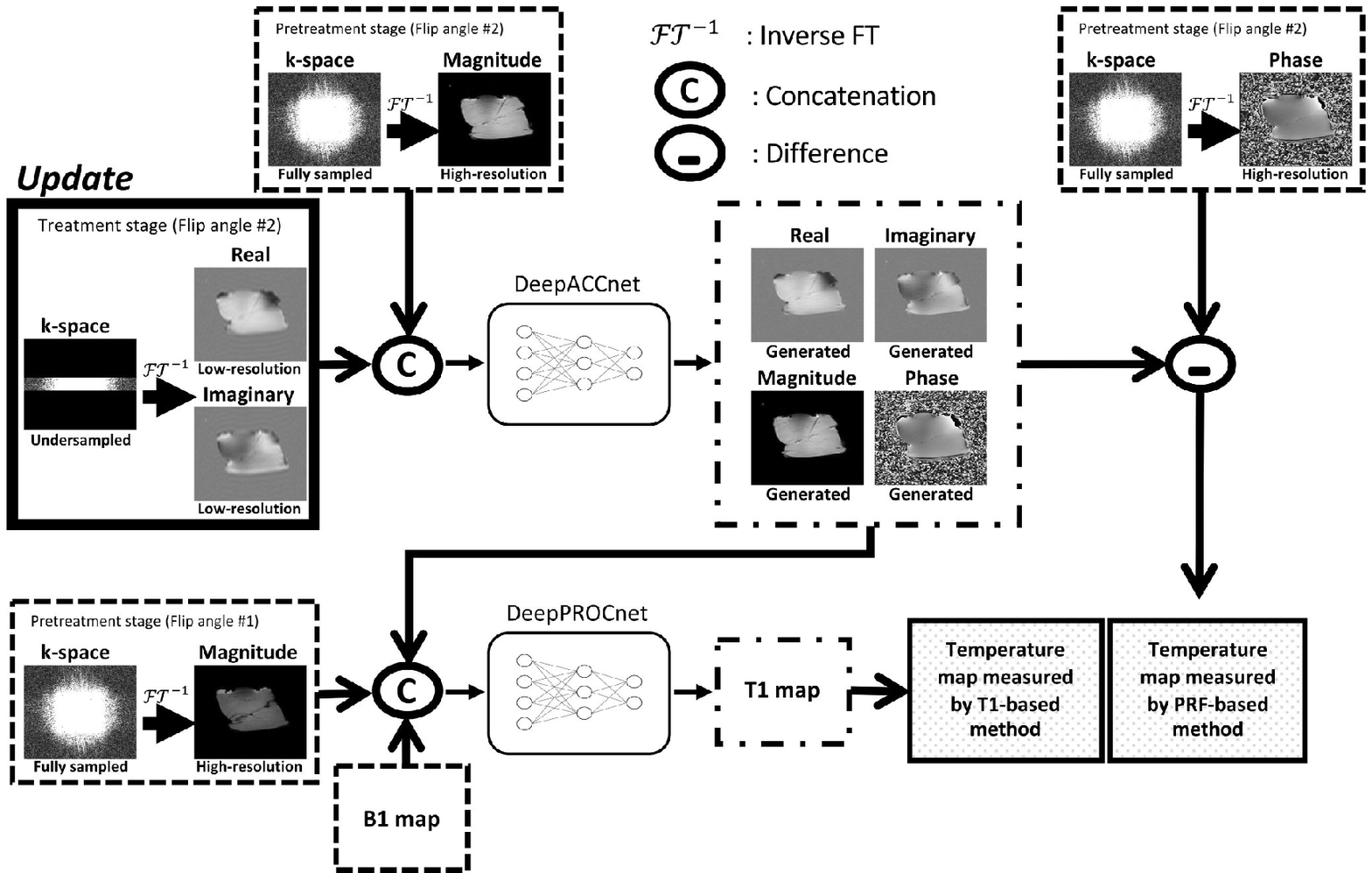}
   \captionv{12}{Short title - can be blank}{Schematic illustration of the proposed method. The proposed cascaded convolutional neural network (CNN) consists of a DeepACCnet for high-resolution (HR) complex MR image reconstruction and a DeepPROCnet for T1 mapping. The DeepACCnet performs the reconstruction from low-resolution (LR) complex MR images to HR complex MR images. The performance of DeepACCnet is increased by incorporating the HR magnitude MR image acquired during the pretreatment stage. Then, the DeepACCnet-generated HR magnitude MR image with higher flip angle (FA) and HR magnitude MR image with lower FA are used as input and T1 with single reference variable flip angle is used as the label for DeepPROCnet. PRF: proton resonance frequency, FT: Fourier transform
 
   \label{Fig2} 
    }  
    \end{center}
\end{figure}

The proposed cascaded CNN architecture consists of the DeepACCnet CNN for HR complex MR image reconstruction and the DeepPROCnet CNN for T1 mapping (Fig. \ref{Fig2}). 

First, the DeepACCnet reconstructs the LR complex MR image to an HR complex MR image. To implement the deep learning approaches that are usually implemented in the real domain, we convert the complex-valued constraint to a real-valued constraint\cite{Han2018_Arxiv}. For this purpose, the operator $\mho$: $\mathbb{C}^{N} \longrightarrow \mathbb{R}^{N \times 2}$ is defined as

\begin{equation}\label{eq_2_B_1}
    \mho[\widehat{z}]:= \pmatrix {Re(\widehat{z}) & Im(\widehat{z})}, \forall{\widehat{z}\in\mathbb{C}^{N}},
\end{equation}\\
where $Re(\cdot)$ and $Im(\cdot)$ denote the real and imaginary parts of the argument, respectively. Similarly, we define its inverse operator $\mho^{-1}$: $\mathbb{R}^{N \times 2} \longrightarrow \mathbb{C}^{N}$ as

\begin{equation}\label{eq_2_B_2}
    \mho^{-1}[\widehat{Z}]:= \widehat{z}_{1}+\imath \widehat{z}_{2}, \forall\widehat{Z}:= \pmatrix{\widehat{z_{1}} & \widehat{z_{2}}}\in\mathbb{R}^{N \times 2}.
\end{equation}\\
Then, the DeepACCNet operator is defined as $\Psi$: $\mathbb{R}^{N_{x} \times N_{y} \times 3} \longrightarrow \mathbb{R}^{N_{x} \times N_{y} \times 2}$. The algorithm used to reconstruct the HR complex MR image is 

\begin{equation}\label{eq_2_B_3}
    \mathcal{Y}_{\mathcal{ACC}}=\mho^{-1}[\Psi(\mathcal{\mathcal{X}_{\mathcal{PRIOR}} \circledast \mho[\mathcal{X}_{\mathcal{ACC}}] })]
\end{equation}
where $\mathcal{X}_{\mathcal{PRIOR}}\in\mathbb{R}^{N_{x} \times N_{y} \times 1}$  is the magnitude of the MR image acquired during the pretreatment stage, $\mathcal{M}(S_{F}(FA_{2}, t_{0}; TE_{l}))$, $\circledast$ is the concatenate in 3-dimensions, $\mathcal{X}_{\mathcal{ACC}}\in\mathbb{C}^{N_{x} \times N_{y} \times 1}$ is the complex-valued MR image acquired during the treatment stage, and $S_{U}(FA_{2}, t_{i}; TE_{l})$. $\mathcal{Y}_{\mathcal{ACC}}\in\mathbb{C}^{N_{x} \times N_{y} \times 1}$ is the DeepACCnet-generated HR complex MR image, $S_{G}(FA_{2}, t_{i}; TE_{l})$.

The second CNN is the DeepPROCnet, which performs the T1 mapping. To increase the flexibility of the network, the input data was split by patches. The patch-split operator can be defined as $\wp$: $\mathbb{R}^{N_{x} \times N_{y}} \longrightarrow \mathbb{R}^{ \frac{N_{x}N_{y}}{N_{p}N_{q}} \times N_{p} \times N_{q}  }$. Similarly, we define its inverse operator, $\wp^{-1}$: $\mathbb{R}^{\frac{N_{x}N_{y}}{N_{p}N_{q}} \times N_{p} \times N_{q}} \longrightarrow \mathbb{R}^{N_{x} \times N_{y}}$. The DeepPROCnet operator is $\Upsilon$: $\mathbb{R}^{\frac{N_{x}N_{y}}{N_{p}N_{q}} \times N_{p} \times N_{q} \times 3} \longrightarrow \mathbb{R}^{\frac{N_{x}N_{y}}{N_{p}N_{q}} \times N_{p} \times N_{q} \times 1}$. Then, the algorithm used to calculate the T1 relaxation time is 
\begin{equation}\label{eq_2_B_4}
    \mathcal{Y}_{\mathcal{PROC}}=\wp^{-1}[\Upsilon(\wp[\mathcal{X}_{\mathcal{PROC}}])],
\end{equation}
where $\mathcal{X}_{\mathcal{PROC}}\in\mathbb{R}^{N_{x} \times N_{y} \times 3}$ is the input of the DeepPROCnet, $\mathcal{M}(S_{F}(FA_{1}, t_{0}; TE_{1})) \circledast \mathcal{M}(S_{G}(FA_{2}, t_{i}; TE_{1}) \circledast B1(t_{0}))$, and $\mathcal{Y}_{\mathcal{PROC}}\in\mathbb{R}^{N_{x} \times N_{y} \times 1}$ is the DeepPROCnet generated T1 map, $T1_{g}(t_{i};TE_{1})$. The T1 change, $\Delta T1(t_{i};TE_{1})$, is calculated as
\begin{equation}\label{eq_2_B_5}
    \Delta T1(t_{i};TE_{1}) = T1_{g}(t_{i};TE_{1}) - T1(t_{0};TE_{1}),
\end{equation}
where $T1(t_{0};TE_{1})$ is the T1 map acquired during the pretreatment stage.

The temperature change map according to the PRF-shift, $\Delta T _{PRF}(t_{i};TE_{l})$, is calculated as
\begin{equation}\label{eq_2_B_6}
    \Delta T _{PRF}(t_{i};TE_{l}) = \frac{\Phi(S_{F}(FA_{2}, t_{0}; TE_{l}) \cdot conj(S_{G}(FA_{2}, t_{i}; TE_{l})))}{\gamma \cdot \alpha \cdot TE_{l} \cdot B_{0} },
\end{equation}
where $\Phi(\cdot)$ and $conj(\cdot)$ are the phase and conjugate of the argument, respectively. $\gamma$, $\alpha$, and $B_{0}$ are the gyromagnetic ratio, the chemical-shift coefficient, which was assumed to be 0.01 $ppm/^{\circ}$C, and the main field strength, respectively. Because PRF-based MR thermometry is a simple process, the temperature change map was directly calculated.

The schematic illustration of the proposed method is shown in Fig. \ref{Fig2}. To train the network, we first apply the inverse Fourier transform to the fully sampled k-space acquired with lower and higher FAs during the pretreatment stage and then apply the inverse Fourier transform to the undersampled k-space, which is filled with zeros, acquired during the treatment stage (higher FA) to generate the LR complex MR image. The region outside the subject was masked to remove random values, which complicate the network training. To acquire the MR image using a multi-channel receiver radio-frequency (RF) coil case, the multi-channel complex MR image was combined by Roemer's method\cite{roemer1990_MRM}. Although the network model for DeepACCnet and DeepPROCnet, which is a cascaded version of U-Net\cite{ronneberger2015u}, has the same structure, the number of input and output channels is different. The numbers of channels of the convolutional layers are summarized at the bottom of the blocks in Supplementary information Fig. \ref{SFigS1}. The input of DeepACCnet is the concatenation of the real and imaginary parts of the LR complex MR image acquired during the treatment stage and the HR magnitude MR image acquired with higher FA during the pretreatment stage. The output of DeepACCnet consists of the real and imaginary parts of the generated HR complex MR image. The input of DeepPROCnet is the concatenation of the DeepACCnet-generated HR magnitude MR image acquired during the treatment stage, the HR magnitude MR image acquired with lower FA during the pretreatment stage, and the B1 map acquired during the pretreatment stage. The output of DeepPROCnet is the generated HR T1 map. Therefore, the number of input channels for DeepACCnet and DeepPROCnet are 3 and 3, respectively, and the number of output channels for DeepACCnet and DeepPROCnet are 2 and 1, respectively. 

\section{Experiments}

\subsection{Data}
The study protocol was approved by the Institutional Review Board of Korea University. The proposed method was evaluated through heating experiments of an agarose gel phantom and ex vivo porcine muscle with and without fat layers, and non-heating experiments of in vivo human prostate and brain. The phantom consisted of 2.0$\%$ agarose gel doped with NaCl (0.5$\%$) and CuSO$_{4}$ (0.1$\%$). 

To train DeepPROCnet, MR images from the knee, brain, and low pelvic region of four healthy volunteers; ex vivo porcine muscle; and agarose gel phantom were used. A 32$\times$32 voxels 2D patch was used for training and testing. For training DeepPROCnet, the patch was generated with an overlapping scheme of 50$\%$ overlap to adjacent patches. DeepPROCnet training data were acquired using meGRE with srVFA and the dual repetition time (TR) method\cite{yarnykh2007MRM}. Detailed parameters for meGRE with srVFA are: field-of-view (FOV) = 224$\times$224 mm$^{2}$, voxel size = 2$\times$2$\times$5 mm$^{3}$, number of slices = 11, number of dynamic scans = 1, TR = 50 ms, TEs = 3, 5.6, 8.2, 10.8, and 13.4 ms, and FA = 10 and 35$^{\circ}$. Detailed parameters for the B1 map are: field-of-view (FOV) = 224$\times$224 mm$^{2}$, voxel size = 2$\times$2$\times$5 mm$^{3}$, number of slices = 11, TRs = 25 and 125 ms, TE = 2 ms, and FA = 60$^{\circ}$. For the in vivo human prostate, MR images were split into 224$\times$224 sections.

To train the DeepACCnet, MR images from the knee, brain, and low pelvic region of four healthy volunteers, the brains of five healthy volunteers, ex vivo porcine muscle, and agarose gel phantom were used. DeepACCnet training data were also acquired using meGRE and the detailed parameters are: FOV = 224$\times$224 mm$^{2}$, voxel size = 2$\times$2$\times$5 mm$^{3}$, number of slices = 11, number of dynamic scans = 60, TR = 50 ms, TEs = 3, 5.6, 8.2, 10.8, and 13.4 ms, and FA = 10 and 35$^{\circ}$. In the in vivo cases, MR images were acquired without high-intensity focused ultrasound (HIFU) sonication and the number of dynamics was 5. In the ex vivo and phantom cases, MR images were acquired with HIFU sonication. 

For the tests, MR images using meGRE with srVFA were acquired with a 3 T MRI scanner (Philips Achieva, the Netherlands) equipped with a HIFU (EofE Ultrasonics, Inc., South Korea, center frequency = 2.31 MHz). The MR parameters were as follows: FOV = 224$\times$224 mm$^{2}$; voxel size = 2$\times$2$\times$5 mm$^{3}$; number of slices = 1; number of dynamic scans = 60; TR = 50 ms; TEs = 3, 5.6, 8.2, 10.8, and 13.4 ms; and FA = 10 and 35$^{\circ}$. For the performance evaluation of DeepACCnet, the acquired MR image was retrospectively undersampled with 16 encoding steps. For the in vivo human prostate, MR images were split into 224$\times$224 sections. To evaluate the precision of MR temperature estimation, the temperature changes of the agarose gel phantom and the ex vivo porcine muscle with and without fat layers were measured using MR-compatible fiber-optic temperature sensors (Photon Control, Inc., Alberta, CA, USA).

\subsection{Network training}
The DeepACCnet model consisted of 19 convolutional layers, 18 batch normalization, 18 rectified linear unit (ReLU) nonlinear layers, 4 max-pooling layers, 4 transposed convolutional layers, and 4 feature contracting paths. The first half of the network consisted of four groups, and each group contained two sets of convolutional layers with a 5$\times$5$\times$5 kernel, batch normalization, and ReLU layer. Each group was connected by a max-pooling layer. The second half of the network also consisted of four groups containing additional feature-concatenation layers compared to the first half’s groups. Each group was connected by a transposed convolution layer instead of a max-pooling layer. Two groups were connected by two convolutional layers. Finally, the last layer applied a 1$\times$1$\times$1 convolution kernel. For the DeepPROCnet, the model used was identical except for the number of channels. For DeepACCnet, minimization was performed with the Adadelta Optimizer. The learning rate was 10$^{-2}$. The batch size was set to 12, and training was stopped after 100 epochs, as the performance at that point became stable. For DeepPROCnet, minimization was also performed with the Adadelta Optimizer. The learning rate was 10$^{-5}$. The batch size was set to 200, and training was also stopped at 100 epochs. The network was trained and evaluated using the Keras software package running on an NVIDIA 1080Ti GPU. The total training time was approximately 92 h.

\subsection{Comparison study}
The proposed DeepACCnet was compared using four methods: (1) fully sampled image; (2) keyhole method; (3) zero filled method; and (4) cascaded CNN without HR acquired during the pretreatment stage (the input of equation~\ref{eq_2_B_3}, where $\mathcal{\mathcal{X}_{\mathcal{PRIOR}} \circledast \mho[\mathcal{X}_{\mathcal{ACC}}] }$
 is replaced with $\mathcal{ \mho[\mathcal{X}_{\mathcal{ACC}}] }$
). The DeepPROCnet is compared with the conventional curve-fitting method. Therefore, six combinations of the algorithms are compared: (1) fully sampled + curve fitting; (2) keyhole + DeepPROCnet; (3) zero filled + DeepPROCnet; (4) cascaded CNN with HR; (5) cascaded CNN without HR; and (6) fully sampled + DeepPROCnet. Fully sampled + curve fitting was considered the ground truth.

\subsection{Data analysis}
Differences between the results of the cascaded CNN generated temperature and T1 map and the ground truth were evaluated across subjects based on the normalized mean-square error ($NMSE$) and structural similarity ($SSIM$). The $NMSE$ for output image $\mathcal{X}$ and ground truth $\mathcal{Y}$ is calculated by
\begin{equation}\label{eq_3_C_1}
    NMSE(\mathcal{X},\mathcal{Y})=\sum \frac{\parallel \mathcal{X}-\mathcal{Y} \parallel^{2}_{2}}{{\parallel \mathcal{Y} \parallel}^{2}_{2}}.
\end{equation}
$SSIM$ is defined by
\begin{equation}\label{eq_3_C_2}
    SSIM(\mathcal{X},\mathcal{Y})=[l(\mathcal{X},\mathcal{Y})^{\alpha} \cdot c(\mathcal{X},\mathcal{Y}))^{\beta} \cdot s(\mathcal{X},\mathcal{Y}))^{\gamma}],
\end{equation}
where the three weights ($\alpha$, $\beta$, and $\gamma$) were set to 1 and the luminance ($l$), contrast ($c$), and structure ($s$) are described in the previous study\cite{Wang2003_multiscale}. To exclude the effects of the background, the center quarter of each image was cropped out and evaluated using the $SSIM$ function provided by MATLAB 2018b (Mathworks, Natick, MA, USA).

\section{Results}

\subsection{Experiment validation by fiber-optic sensor}

\begin{figure}[ht]
   \begin{center}
   \includegraphics[width=14cm]{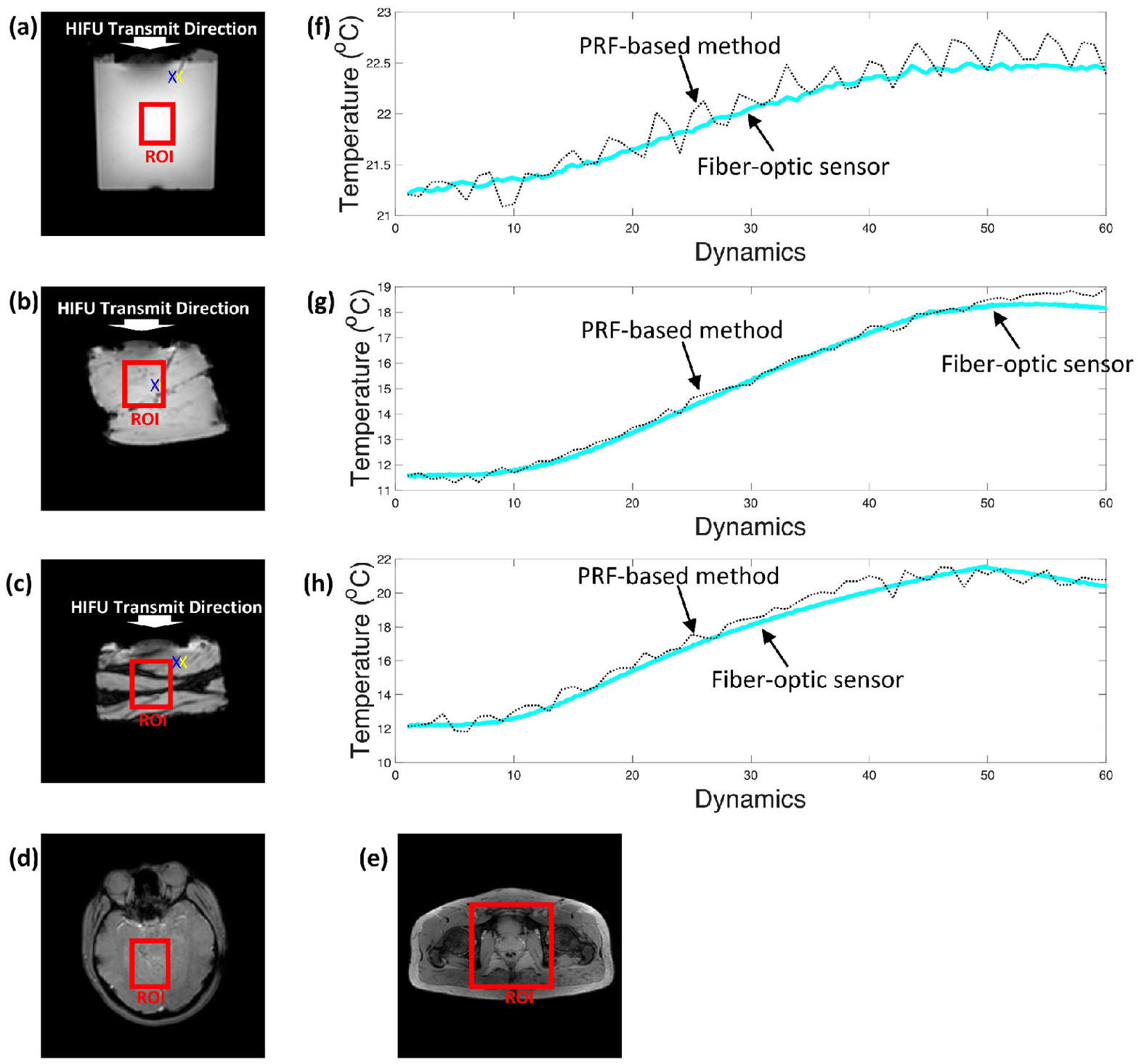}

   %
   %
   \captionv{12}{Short title - can be blank}{Axial magnitude MR images and experimental setup for (a) the agarose gel phantom; (c) porcine muscle; (e) porcine muscle with fat layers: (g) in vivo human brain; and (h) in vivo human prostate. The time through temperature change of fiber-optic sensors and the PRF-based method for (b) the agarose gel phantom; (d) porcine muscle; and (f) porcine muscle with fat layers. HIFU: high-intensity focused ultrasound, ROI: region-of-intest, PRF: proton resonance frequency 
   \label{Fig3} 
    }  
    \end{center}
\end{figure}

The axial magnitude MR images and experimental setup for the agarose gel phantom, ex vivo porcine muscle, ex vivo porcine muscle with fat layers, in vivo human brain, and in vivo human prostate are shown in Fig. \ref{Fig3}. For comparison with a fiber-optic temperature sensor, temperatures measured by PRF-based MR thermometry were averaged over a 5$\times$5-pixel region of interest (ROI), with the center located adjacent to the fiber-optic temperature probe. The temperatures from the cascaded CNN with HR and fiber-optic probe measurements are plotted in Figs. \ref{Fig3}f--h. Based on 60 dynamic scans, the maximum differences between the cascaded CNN with HR and the fiber-optic sensor for the agarose gel phantom, ex vivo porcine muscle, and ex vivo porcine muscle with fat layers were 0.32, 0.58, and 0.94 $^{\circ}$C, respectively. The location and size of their ROIs, plus those of in vivo human brain and in vivo human prostate for Bland-Altman analysis, and the corresponding $SSIM$ and $NMSE$ values are shown in Fig. \ref{Fig3}. The sizes of the ROIs are 20$\times$25, 40$\times$20, 20$\times$25, 50$\times$40, and 150$\times$140, respectively.

\subsection{Model validation by uniform agarose gel phantom and porcine muscle heating experiments}

\begin{figure}[ht]
   \begin{center}
   \includegraphics[width=14cm]{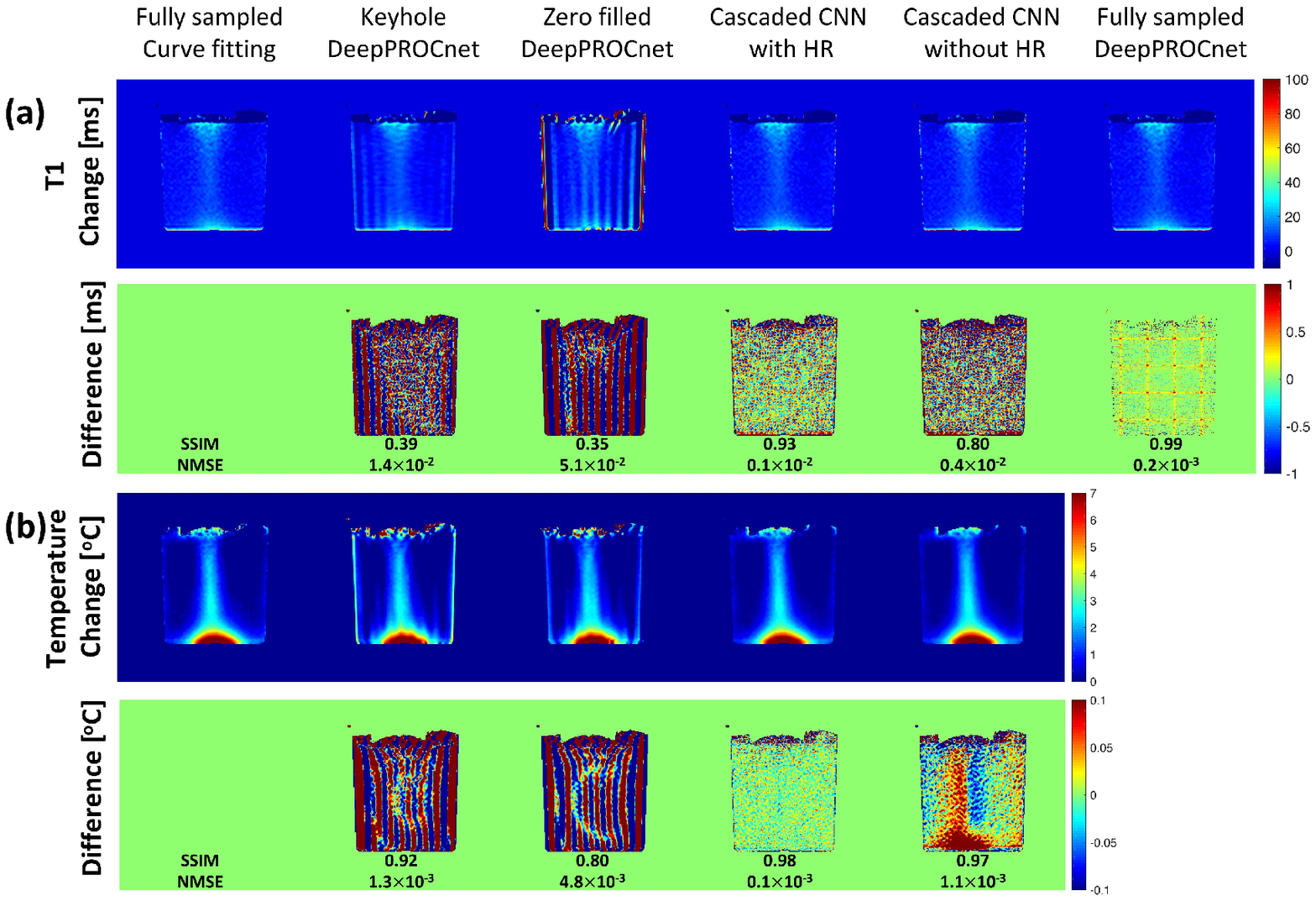}

   %
   %
   \captionv{12}{Short title - can be blank}{Agarose gel phantom heating results. (a) T1 change maps reconstructed with fully sampled + curve fitting, keyhole + DeepPROCnet, zero filled + DeepPROCnet, cascaded convolutional network (CNN) with high resolution (HR), cascaded CNN without HR, and fully sampled + DeepPROCnet. (b) Temperature change maps reconstructed with fully sampled + curve fitting, keyhole + DeepPROCnet, zero filled + DeepPROCnet, cascaded convolutional network (CNN) with high resolution (HR), and cascaded CNN without HR. $SSIM$ and $NMSE$ values are shown below each image. CNN: convolutional neural network, HR: high-resolution, $SSIM$: structural similarity, $NMSE$: normalized mean-square error
   \label{Fig4} 
    }  
    \end{center}
\end{figure}

Figure \ref{Fig4}a shows the T1 change maps at the 50th dynamic scan, reconstructed with fully sampled + curve fitting, keyhole + DeepPROCnet, zero filled + DeepPROCnet, cascaded CNN with HR, cascaded CNN without HR, and fully sampled + DeepPROCnet, respectively. When the cascaded CNN is not used, an aliasing artifact can be clearly observed. In addition, the $SSIM$/$NMSE$ values when using the cascaded CNN were significantly better than those without the cascaded CNN (0.93/0.1$\times$10$^{-2}$ for the cascaded CNN with HR and 0.80/0.4$\times$10$^{-2}$ for the cascaded CNN without HR). As shown by the differences in the T1 change maps calculated through fully sampled + curve fitting and fully sampled + DeepPROCnet, the $SSIM$ and $NMSE$ values for DeepPROCnet are 0.99 and 0.2$\times$10$^{-3}$, respectively. Figure \ref{Fig4}b shows the temperature change maps measured by PRF-based MR thermometry at the 50th dynamic scan, reconstructed with fully sampled + curve fitting, keyhole + DeepPROCnet, zero filled + DeepPROCnet, cascaded CNN with HR, and cascaded CNN without HR. When the cascaded CNN was not used, the aliasing artifact was clearly visible. The $SSIM$/$NMSE$ values when using a cascaded CNN were significantly better than those without the cascaded CNN (0.98/0.1$\times$10$^{-3}$ for the cascaded CNN with HR and 0.97/1.1$\times$10$^{-3}$ for the cascaded CNN without HR). The cascaded CNN with HR shows that the shape of the hot spot has not changed, either in the T1 or temperature change images. However, when the HR image is not used, the shape of the hot spot changes slightly.

\begin{figure}[ht]
   \begin{center}
   \includegraphics[width=14cm]{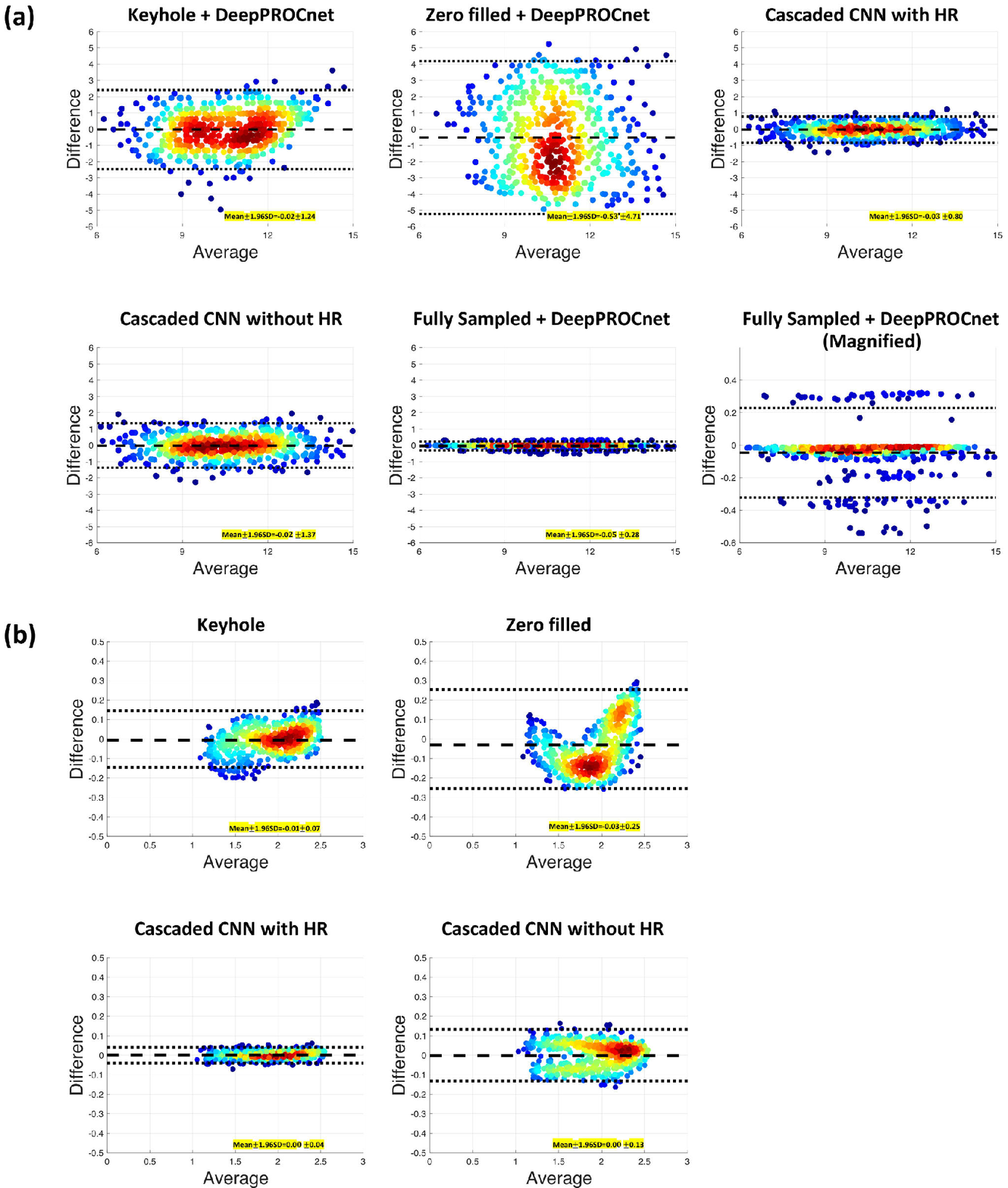}

   %
   %
   \captionv{12}{Short title - can be blank}{Bland-Altman results in agarose gel phantom for T1 and proton resonance frequency-based temperature change reconstructed with fully sampled + curve fitting, keyhole + DeepPROCnet, zero filled + DeepPROCnet, cascaded convolutional network (CNN) with high resolution (HR), cascaded CNN without HR, and fully sampled + DeepPROCnet. Bias and limit of agreement values are shown below each image. CNN: convolutional neural network, HR: high-resolution
   \label{Fig5} 
    }  
    \end{center}
\end{figure}

Side-by-side comparisons of the T1 change and temperature change between ground truth and keyhole + DeepPROCnet, zero filled + DeepPROCnet, cascaded CNN with HR, cascaded CNN without HR, and fully sampled + DeepPROCnet are shown in Fig. \ref{Fig5}. T1 and temperature change images reconstructed by cascaded CNN with HR have smaller limits of agreement. The bias and limit of agreement for the T1 change with cascaded CNN with HR are -0.03 and 0.80, respectively. Those for temperature change with cascaded CNN with HR are 0.00 and 0.04, respectively. Most errors in DeepPROCnet are caused by the patch-splitting performed in the data augmentation step. In addition, by incorporating the HR, the limit of agreement showed a decreasing tendency in the case of the T1 change, but the bias was calibrated in hotspots for PRF-based MR thermometry.

\begin{figure}[ht]
   \begin{center}
   \includegraphics[width=14cm]{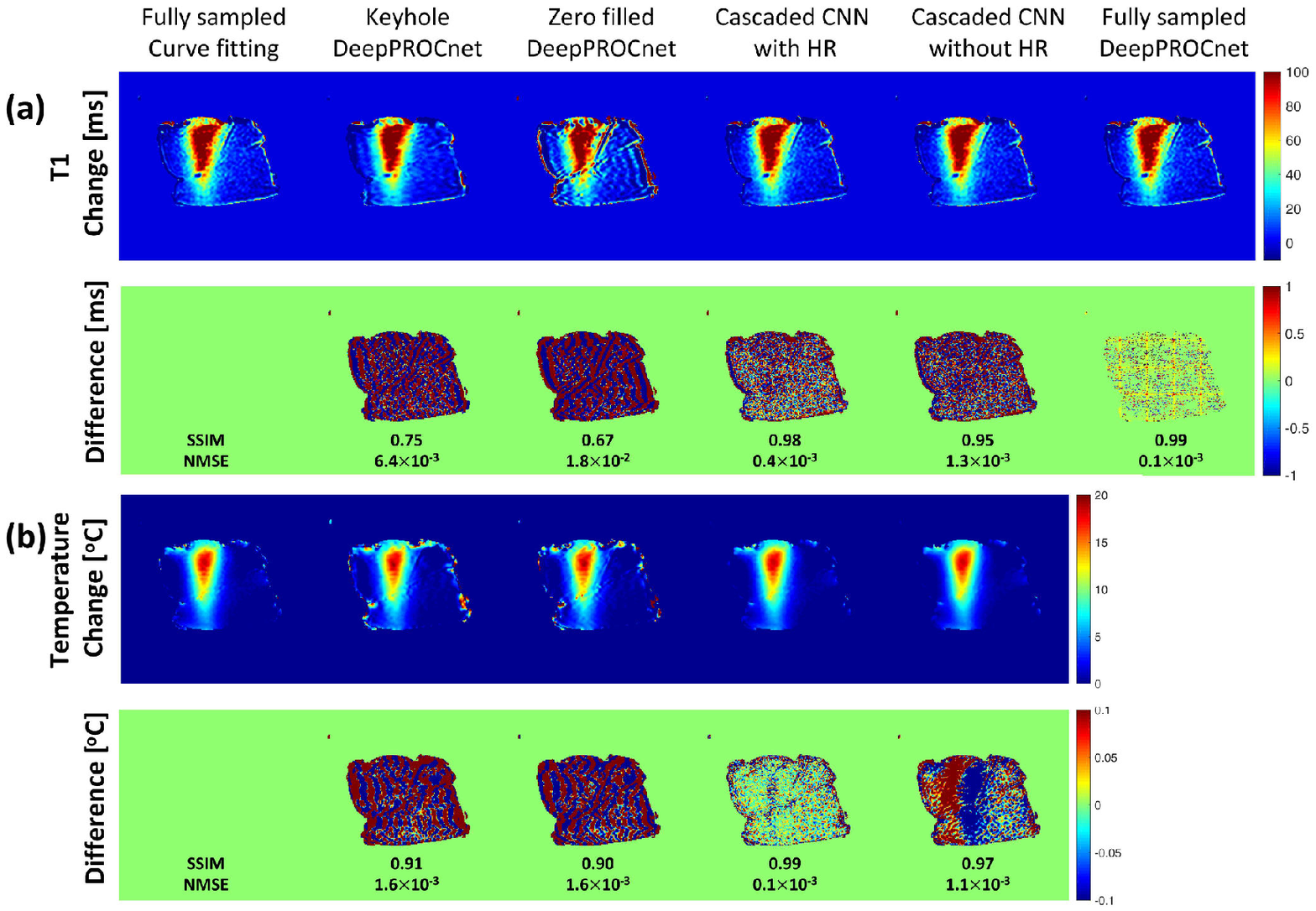}
   \captionv{12}{Short title - can be blank}{Ex vivo porcine muscle heating results. (a) T1 change maps reconstructed with fully sampled + curve fitting, keyhole + DeepPROCnet, zero filled + DeepPROCnet, cascaded convolutional network (CNN) with high resolution (HR), cascaded CNN without HR, and fully sampled + DeepPROCnet. (b) Temperature change maps reconstructed with fully sampled + curve fitting, keyhole + DeepPROCnet, zero filled + DeepPROCnet, cascaded convolutional network (CNN) with high resolution (HR), and cascaded CNN without HR. $SSIM$ and $NMSE$ values are shown below each image. CNN: convolutional neural network, HR: high-resolution, $SSIM$: structural similarity, $NMSE$: normalized mean-square error
   \label{Fig6} 
    }  
    \end{center}
\end{figure}

T1 and temperature change maps in ex vivo porcine muscle during HIFU sonication are shown in Fig. \ref{Fig6} at the time of the peak temperature rise. Time-lapse videos of the sonication taking place in Fig. \ref{Fig6} are shown in Supporting information video 1. Figure \ref{Fig6}a shows the T1 change maps reconstructed with fully sampled + curve fitting, keyhole + DeepPROCnet, zero filled + DeepPROCnet, cascaded CNN with HR, cascaded CNN without HR, and fully sampled + DeepPROCnet, respectively. The $SSIM$/$NMSE$ values when using the cascaded CNN were significantly better than those without the cascaded CNN (0.98/0.4$\times$10$^{-3}$ for the cascaded CNN with HR and 0.95/1.3$\times$10$^{-3}$ for the cascaded CNN without HR). Figure \ref{Fig6}b shows the temperature change maps at the time of the peak temperature rise, reconstructed with fully sampled + curve fitting, keyhole + DeepPROCnet, zero filled + DeepPROCnet, cascaded CNN with HR, and cascaded CNN without HR, respectively. When using cascaded CNN, the $SSIM$/$NMSE$ values were significantly better (0.99/0.1$\times$10$^{-3}$ for the cascaded CNN with HR and 0.97/1.1$\times$10$^{-3}$ for the cascaded CNN without HR). Using cascaded CNN with HR does not change the shape of the hot spot in either the T1 or temperature change images.

\begin{figure}[ht]
   \begin{center}
   \includegraphics[width=14cm]{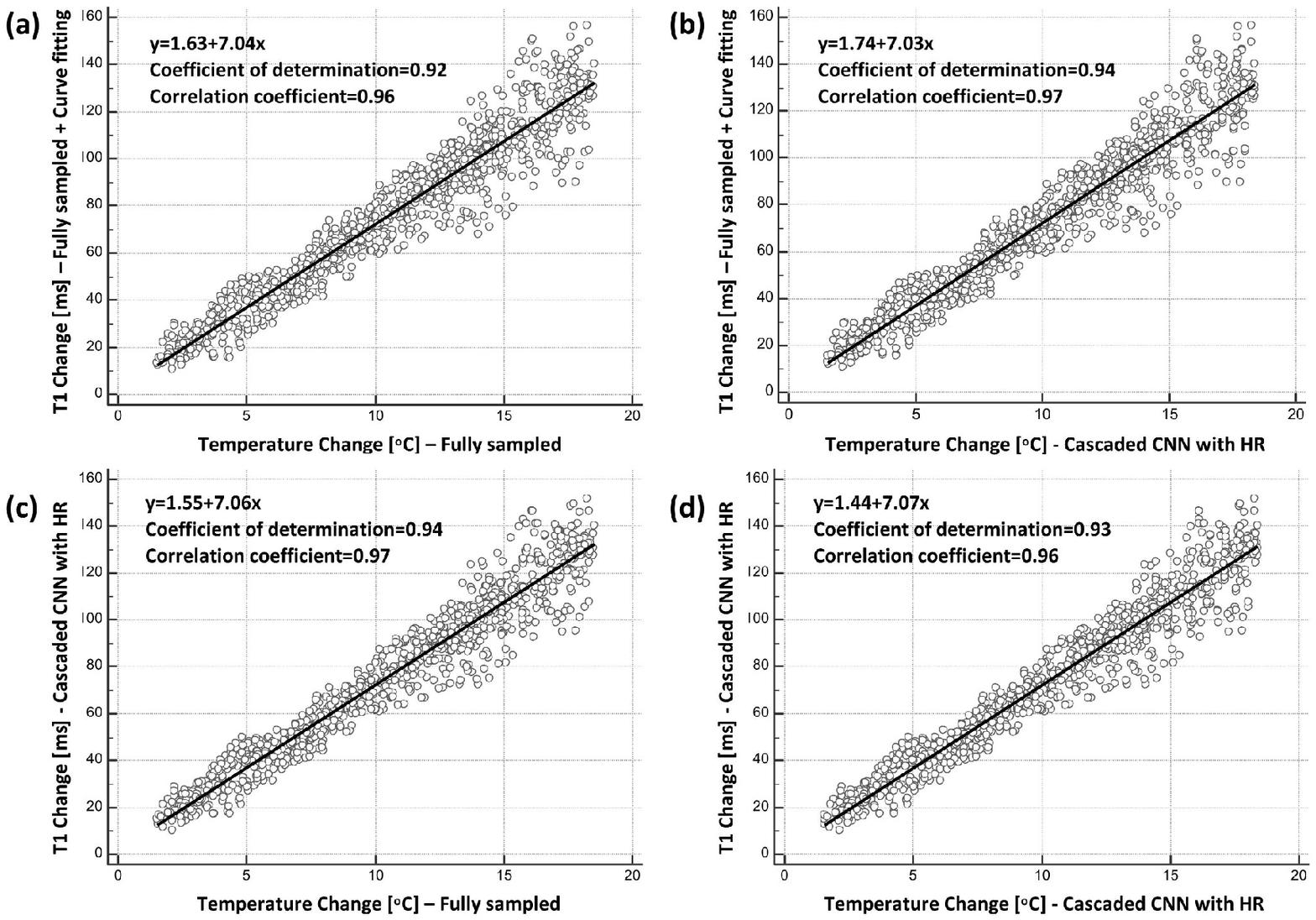}
   \captionv{12}{Short title - can be blank}{Temperature change by proton resonance freqeuncy-based MR thermometry versus T1 change for ex vivo porcine muscle. (a) Relationship between temperature change by the fully sampled and T1 change by fully sampled + curve fitting. (b) Relationship between temperature change by the cascaded convolutional neural network (CNN) with high resolution (HR) and T1 change by fully sampled + curve fitting. (c) Relationship between temperature change by the fully sampled and T1 change by cascaded CNN with HR. (d) Relationship between temperature change by the cascaded CNN with HR and T1 change by the cascaded CNN with HR. CNN: convolutional neural network, HR: high resolution
   \label{Fig7} 
    }  
    \end{center}
\end{figure}

Figure \ref{Fig7} provides a point by point comparison of the T1 change versus temperature change for the ex vivo porcine muscle. The slope from a linear fit to the data was 7.04 ms/$^{\circ}$C (the coefficient of determination is 0.92) for T1 changes by fully sampled + curve fitting versus temperature change by fully sampled (Fig. \ref{Fig7}a). The linear fit slope to the data was 7.03 ms/$^{\circ}$C (the coefficient of determination is 0.94) for T1 changes by cascaded CNN with HR versus temperature change by cascaded CNN with HR (Fig. \ref{Fig7}b). The linear fit slope to the data was 7.07 ms/$^{\circ}$C (the coefficient of determination is 0.93) for T1 changes by cascaded CNN with HR versus temperature change by fully sampled (Fig. \ref{Fig7}c). The linear fit slope to the data was 7.07 ms/$^{\circ}$C (the coefficient of determination is 0.93) for T1 changes by cascaded CNN with HR versus temperature change by cascaded CNN with HR (Fig. \ref{Fig7}d).  Although the temperature of T1 depends on the tissue type, the figure shows that the temperature change and T1 change have significantly higher correlation. In addition, when comparing Fig. \ref{Fig7}b and Fig. \ref{Fig7}c, it can be seen that temperature dependency of the proposed method and the temperature dependency of T1 of the ground truth are very similar.

\subsection{Effects of adipose tissue: heating experiment for ex vivo porcine muscle with fat layers}

\begin{figure}[ht]
   \begin{center}
   \includegraphics[width=14cm]{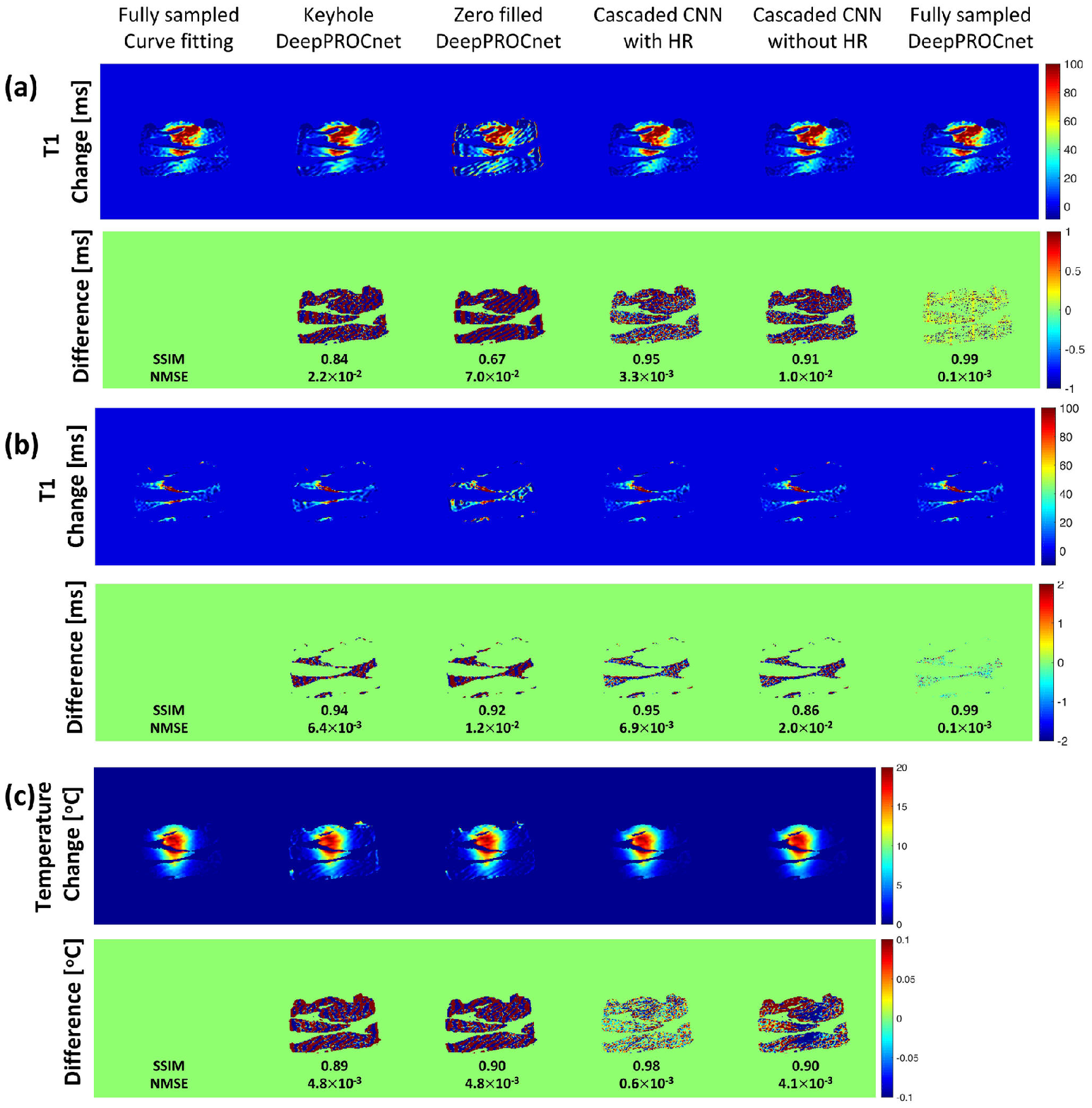}
   \captionv{12}{Short title - can be blank}{Heating results for ex vivo porcine muscle with fat layers. (a) T1 change maps reconstructed with fully sampled + curve fitting, keyhole + DeepPROCnet, zero filled + DeepPROCnet, cascaded convolutional network (CNN) with high resolution (HR), cascaded CNN without HR, and fully sampled + DeepPROCnet. (b) Temperature change maps reconstructed with fully sampled + curve fitting, keyhole + DeepPROCnet, zero filled + DeepPROCnet, cascaded convolutional network (CNN) with high resolution (HR), and cascaded CNN without HR. $SSIM$ and $NMSE$ values are shown below each image. CNN: convolutional neural network, HR: high-resolution, $SSIM$: structural similarity, $NMSE$: normalized mean-square error
   \label{Fig8}
    }  
    \end{center}
\end{figure}

Figure \ref{Fig8}a shows the T1 change maps in aqueous tissue reconstructed with fully sampled + curve fitting, keyhole + DeepPROCnet, zero filled + DeepPROCnet, cascaded CNN with HR, cascaded CNN without HR, and fully sampled + DeepPROCnet, respectively. The $SSIM$/$NMSE$ values when using the cascaded CNN were significantly better than those without cascaded CNN (0.95/3.3$\times$10$^{-3}$ for the cascaded CNN with HR and 0.91/1.0$\times$10$^{-2}$ for the cascaded CNN without HR). Figure \ref{Fig8}b shows the T1 change maps in adipose tissue reconstructed with fully sampled + curve fitting, keyhole + DeepPROCnet, zero filled + DeepPROCnet, cascaded CNN with HR, cascaded CNN without HR, and fully sampled + DeepPROCnet, respectively. The $SSIM$/$NMSE$ values when using the cascaded CNN were significantly better than those without the cascaded CNN (0.95/6.9$\times$10$^{-3}$ for the cascaded CNN with HR and 0.86/2.0$\times$10$^{-2}$ for the cascaded CNN without HR). Figure \ref{Fig8}c shows the temperature change maps reconstructed with fully sampled + curve fitting, keyhole + DeepPROCnet, zero filled + DeepPROCnet, cascaded CNN with HR, and cascaded CNN without HR, respectively. The $SSIM$/$NMSE$ values when using the cascaded CNN were significantly better than those when not using the cascaded CNN (0.98/0.6$\times$10$^{-3}$ for the cascaded CNN with HR and 0.90/4.1$\times$10$^{-3}$ for the cascaded CNN without HR). Time-lapse videos of the sonication in Fig. \ref{Fig8} are shown in the Supporting information video 2.

\begin{figure}[ht]
   \begin{center}
   \includegraphics[width=14cm]{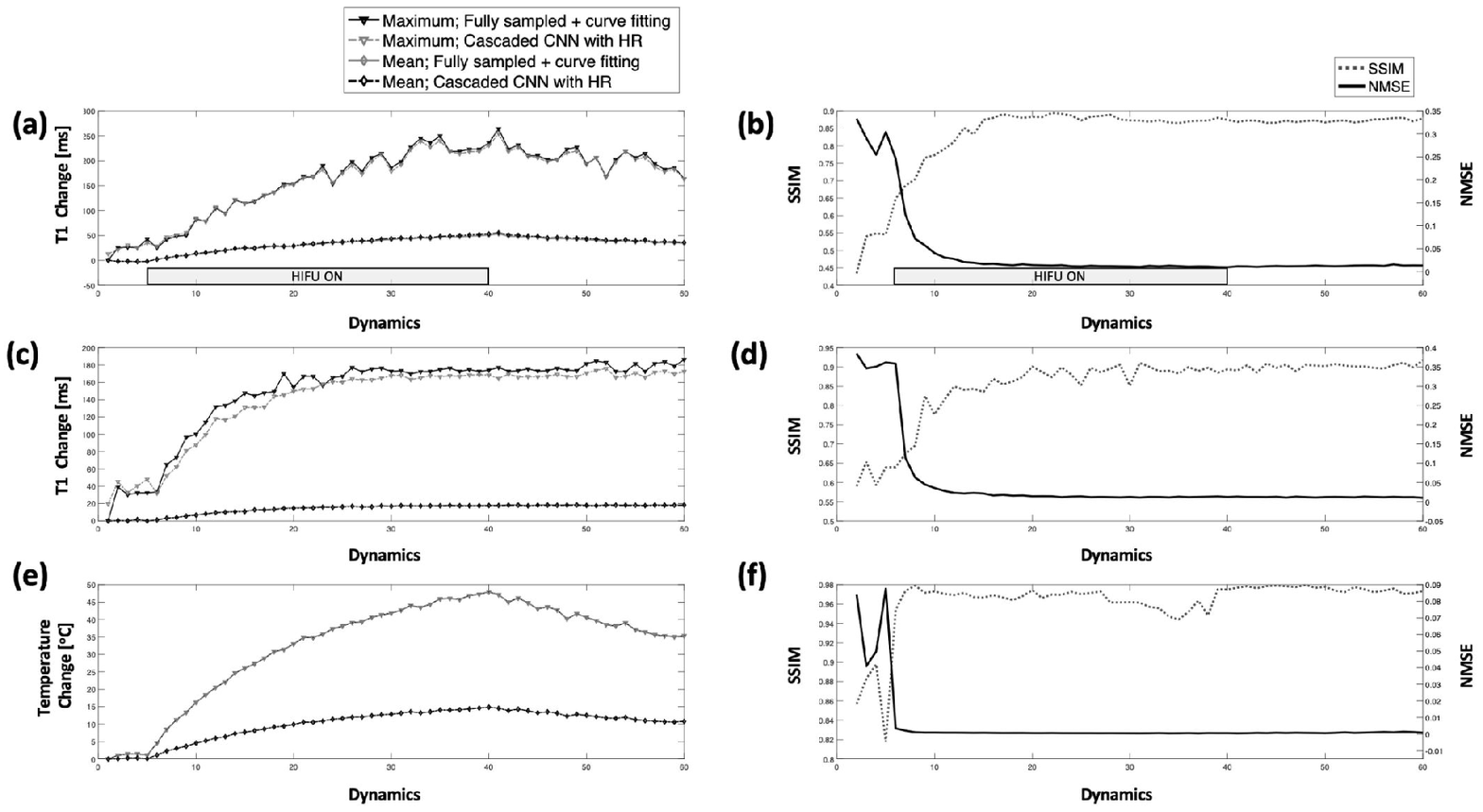}
   \captionv{12}{Short title - can be blank}{Timing results for the ex vivo porcine muscle with fat layers. The time through change of mean and maximum (a); and $SSIM$ and $NMSE$ (d) for T1 change in aqueous tissue. The time through change of mean and maximum (b); and $SSIM$ and $NMSE$ (e) for T1 change in adipose tissue. The time through change of mean and maximum (c); and $SSIM$ and $NMSE$ (f) for temperature change. $SSIM$: structural similarity, $NMSE$: normalized mean-square error
   \label{Fig9} 
    }  
    \end{center}
\end{figure}

The through-time mean, maximum, $SSIM$, and $NMSE$ are shown in Fig. \ref{Fig9}. The temperature and T1 changes show strong agreement in the thermal evolution curves in Figs. \ref{Fig9}a, c, and e. However, in adipose tissues, the maximum T1 change reconstructed by the cascaded CNN with HR is slightly lower than that by the fully sampled + curve fitting method. In the case of adipose tissue, the $SSIM$ values for T1 change fall between 0.9 and 0.95 and the $NMSE$ values for T1 change fall between 4.0$\times$10$^{-2}$ and 3.0$\times$10$^{-3}$. In the case of the aqueous tissue, the $SSIM$ values for T1 change lie between 0.85 and 0.95 and the $NMSE$ values for T1 change lie between 9.0$\times$10$^{-2}$ and 3.0$\times$10$^{-3}$. The $SSIM$ values for temperature change lie between 0.94 and 0.98, and the $NMSE$ values for temperature change measured by PRF-based MR thermometry lie between 0.7$\times$10$^{-3}$ and 0.5$\times$10$^{-3}$.

\subsection{Performance regarding in vivo human datasets}
\begin{figure}[ht]
   \begin{center}
   \includegraphics[width=14cm]{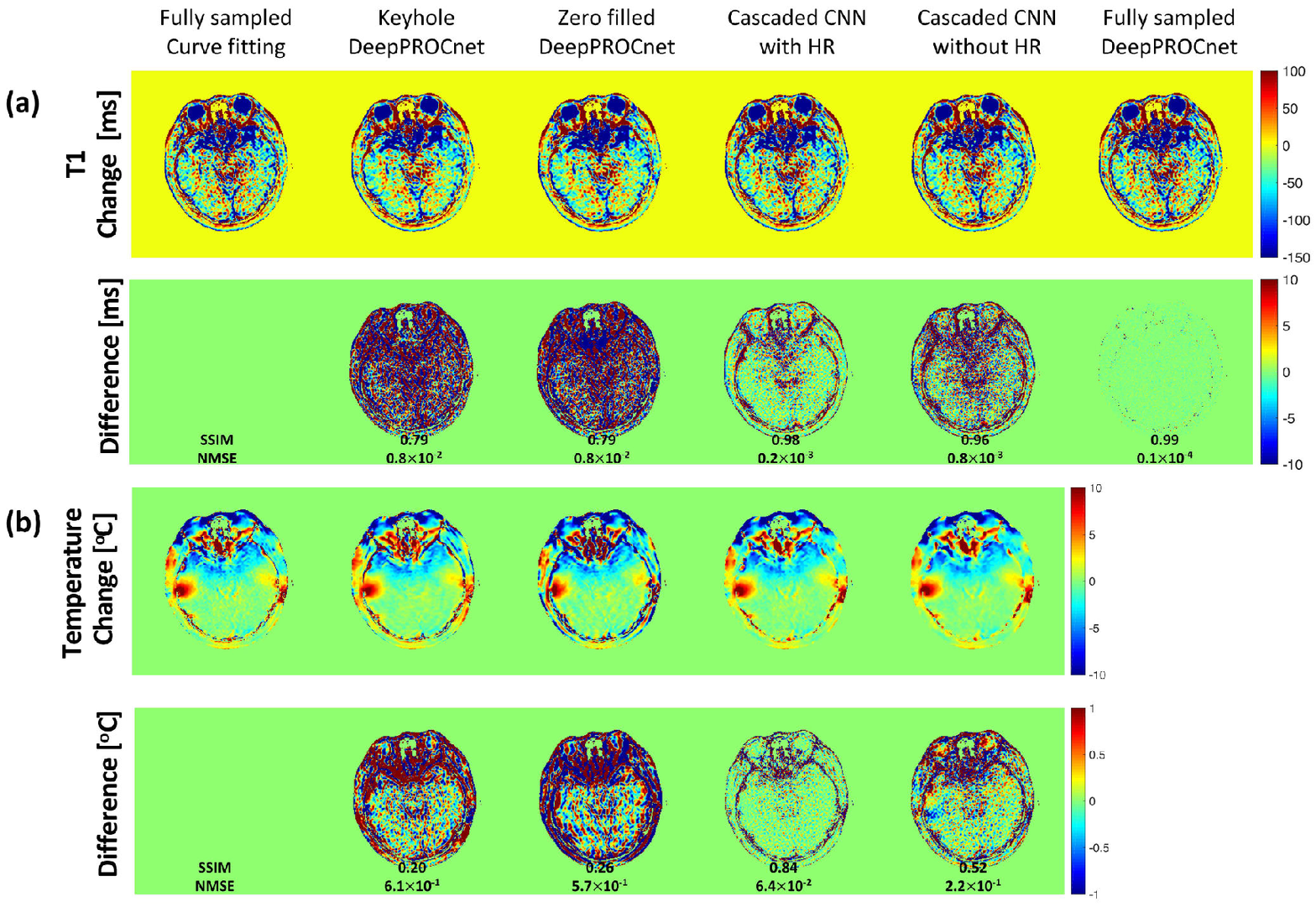}
   \captionv{12}{Short title - can be blank}{In vivo human brain non-heating results. (a) T1 change maps reconstructed with fully sampled + curve fitting, keyhole + DeepPROCnet, zero filled + DeepPROCnet, cascaded convolutional network (CNN) with high resolution (HR), cascaded CNN without HR, and fully sampled + DeepPROCnet. (b) Temperature change maps reconstructed with fully sampled + curve fitting, keyhole + DeepPROCnet, zero filled + DeepPROCnet, cascaded convolutional network (CNN) with high resolution (HR), and cascaded CNN without HR. $SSIM$ and $NMSE$ values are shown below each image. CNN: convolutional neural network, HR: high resolution, $SSIM$: structural similarity, $NMSE$: normalized mean-square error
   \label{Fig10} 
    }  
    \end{center}
\end{figure}

\begin{figure}[ht]
   \begin{center}
   \includegraphics[width=14cm]{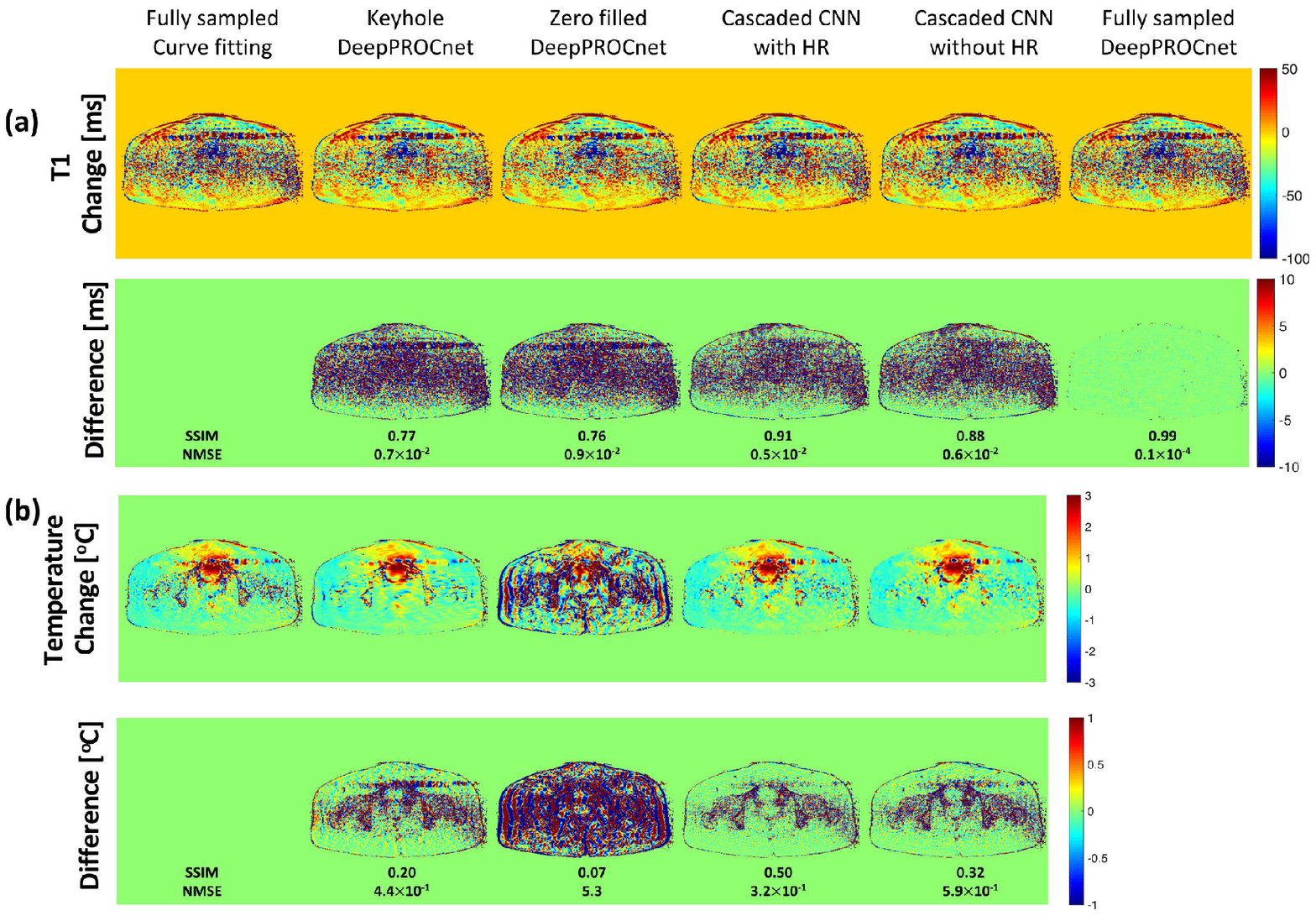}
   \captionv{12}{Short title - can be blank}{In vivo human prostate non-heating results. (a) T1 change maps reconstructed with fully sampled + curve fitting, keyhole + DeepPROCnet, zero filled + DeepPROCnet, cascaded convolutional network (CNN) with high resolution (HR), cascaded CNN without HR, and fully sampled + DeepPROCnet. (b) Temperature change maps reconstructed with fully sampled + curve fitting, keyhole + DeepPROCnet, zero filled + DeepPROCnet, cascaded convolutional network (CNN) with high resolution (HR), and cascaded CNN without HR. $SSIM$ and $NMSE$ values are shown below each image. CNN: convolutional neural network, HR: high resolution, $SSIM$: structural similarity, $NMSE$: normalized mean-square error
   \label{Fig11} 
    }  
    \end{center}
\end{figure}

The in vivo human brain results are shown in Fig. \ref{Fig10}. The $SSIM$/$NMSE$ values for T1 changes were significantly good (0.98/0.2$\times$10$^{-3}$ for the cascaded CNN with HR and 0.96/0.8$\times$10$^{-3}$ for the cascaded CNN without HR). The $SSIM$/$NMSE$ values for temperature change measured by PRF-based MR thermometry were also significantly good (0.84/6.4$\times$10$^{-2}$ for the cascaded CNN with HR and 0.52/2.2$\times$10$^{-1}$ for the cascaded CNN without HR). Especially, assuming that the phase change due to the B0 variation in the ear region is the temperature change, we can observe that the shape and size of the hotspot has changed in the cascaded CNN without HR. In contrast, the shape and size of the hotspot has not changed in the cascaded CNN with HR. The in vivo human prostate results are shown in Fig. \ref{Fig11}. The $SSIM$/$NMSE$ values for T1 changes were significantly better when the cascaded CNN was present (0.91/0.5$\times$10$^{-2}$ for the cascaded CNN with HR and 0.88/0.6$\times$10$^{-2}$ for the cascaded CNN without HR). The $SSIM$/$NMSE$ values for temperature change measured by PRF-based MR thermometry indicated better results when the cascaded CNN was present  (0.50/3.2$\times$10$^{-1}$ for the cascaded CNN with HR and 0.32/5.9$\times$10$^{-1}$ for the cascaded CNN without HR). In addition, the flow artifact was reduced.

\subsection{Processing times of cascaded convolutional neural network}
The computation times needed to acquire a temperature map for simultaneously measuring the aqueous and adipose tissues are shown in Table \ref{Table_1}. Fully sampled + curve fitting and cascaded CNN with HR required 8 h and 32 ms, respectively. In the case of the in vivo human prostate, fully sampled + curve fitting and cascaded CNN with HR required 11 h and 78 ms, respectively.

\section{Discussion}

We have proposed a cascaded CNN that allows processing of real-time interactive MR temperature images in both adipose and aqueous tissues with high image quality. This is made possible by taking advantage of the srVFA, meGRE, cascaded CNN, and data acquisition scheme for MR-guided FUS treatment. During the pretreatment stage, an HR MR image is acquired by meGRE with two FAs. After starting the treatment, a LR MR image is acquired by meGRE with a single FA, which allows for accelerated MR data acquisition during the treatment stage. The MR images acquired in the treatment and pretreatment stages are reconstructed into a T1 map and a temperature map via the cascaded CNN. The first CNN generates a HR complex MR image from the LR complex MR image acquired during the treatment stage, and its performance is enhanced by incorporating the HR magnitude MR image acquired during the pretreatment stage. The second CNN generates the T1 map. The accuracy of the cascaded CNN was confirmed in an agarose gel phantom, ex vivo porcine muscle, ex vivo porcine muscle with fat layers, in vivo human brain, and in vivo human prostate. The cascaded CNN computed temperature maps for the agarose gel phantom, ex vivo porcine muscle, ex vivo porcine muscle with fat layers, in vivo human brain, and in vivo human prostate could be updated every 32 ms, 32 ms, 32 ms, 32 ms, and 78 ms, respectively. The $SSIM$/$NMSE$ values of the T1 change by the cascaded CNN with HR for the agarose gel phantom, ex vivo porcine muscle, ex vivo porcine muscle with fat layers, in vivo human brain, and in vivo human prostate in aqueous tissue are 0.93/0.4$\times$10$^{-2}$, 0.98/0.4$\times$10$^{-3}$, 0.95/3.3$\times$10$^{-3}$, 0.99/0.5$\times$10$^{-3}$, and 0.97/4.1$\times$10$^{-3}$, respectively. The $SSIM$/$NMSE$ values of the T1 change by cascaded CNN with HR for ex vivo porcine muscle with fat layers in adipose tissue are 0.95/6.9$\times$10$^{-3}$. The $SSIM$/$NMSE$ values of the temperature change by cascaded CNN with HR for agarose gel phantom, ex vivo porcine muscle, ex vivo porcine muscle with fat layers, in vivo human brain, and in vivo human prostate are 0.98/0.1$\times$10$^{-3}$, 0.99/0.1$\times$10$^{-3}$, 0.98/0.6$\times$10$^{-3}$, 0.84/6.4$\times$10$^{-2}$, and 0.50/3.2$\times$10$^{-1}$, respectively. 

The magnitude MR image acquired during the pretreatment stage was used to train the DeepACCnet CNN. There are a number of other effects that can influence temperature measurements based on PRF-based MR thermometry, such as the composition of the tissue, magnetic susceptibility, electrical conductivity, and external field drift$\cite{Rieke2013_JMRI}$. Among these, the phase drift induced by the external field drift has a considerable effect on the accuracy of the temperature image and is the main cause behind incorrect temperature images. Because this causes several problems during training, magnitude MR images are considered to be more efficient.

The meGRE sequence is optimal for measuring the T1 and PRF changes because the optimal TEs for VFA and PRF-shift are the shortest one, and when TE is the same as T2*, respectively\cite{Todd2014_MRM}. In addition, because meGRE has a high bandwidth and duty cycle, an image with long TE can be acquired\cite{odeen2019_LSM}. Furthermore, meGRE can be used to reconstruct a material’s magnetic susceptibility and electric conductivity\cite{Wang2015_MRM}\cite{KimJM2018_ISMRM}\cite{Chung2018_ISMRM}. It is expected that the progress of the treatment will be monitored through the magnetic susceptibility and electrical conductivity values. During HIFU sonication, if there is an area in which the attenuation (a scattering property), suddenly changes, a high temperature  is produced, often causing an adverse effect on treatment. In particular, in the case of a prostate or brain, there is often calcification, and characteristic changes caused by calcification affect the temperature rise induced by HIFU sonication, which can lead to failure of the treatment\cite{Barkin2011_CJU}. In addition, it is thought that the strong magnetic susceptibility effect of meGRE allows for monitoring of sudden hemorrhage, among other complications, during treatment\cite{Liang1999_AJN}. 

\subsection{Limitations and future work}

The primary limitation of the proposed method is the potential for deformation and motion of the subject during the treatment stage. This is a problem common to most MR temperature mapping methods. In the case of PRF-based MR thermometry, the solution proposed to address the problems resulting from deformation or motion is to use multi-baseline, referenceless, and fat-referenced PRF-based MR thermometry\cite{Hofstetter2012_JMRI}\cite{Rieke2012_MRM}\cite{Grissom2010_MedPhys}. The multi-baseline approach is a method applied to obtain a baseline library, which is expected to be directly applicable by obtaining multiple baselines in the pretreatment stage from the method proposed herein. In addition, for the referenceless approach, it is likely that PRF-based MR thermometry will be applicable during the reconstruction process, and the fat-referenced approach will also be applicable in the same manner. However, there is a need to consider whether additional processing time, such as polynomial fitting performed in the referenceless or fat-referenced approaches, is required for real-time interactive MR thermometry. In practice, predicting all motion accurately is not possible, and in the case of T1-based MR thermometry, there are currently limitations to applying the referenceless method.

Efforts to understand the characteristics of the deep neural network and explore its benefits for real-time interactive MR temperature imaging will continue. MR-guided surgery requires fast acquisition and processing times for real-time interactive MR temperature imaging, as discussed in this paper. MR images acquired in the pretreatment stage can be used for various purposes, even if the acquisition time is long. In this paper, images were used in two ways, namely to obtain the T1 values using srVFA, and to improve the acquisition time by exploiting the fact that the high frequency components of the MR images do not change significantly. Increasing the resolution of MR images using the HR MR image acquired during the treatment stage is expected to be used in various ways in deep neural networks (e.g., for a discriminator CNN\cite{Kim2018_MedPhys}). In addition, the issues associated with motion or deformation are also expected to diminish by exploiting deep neural networks.

In this study, we used a single-channel coil for signal reception. Although acquiring MR images using multiple channels is expected to achieve better results in DeepACCnet, single-channel reception RF coils are often used to avoid problems such as the size of the water bath or HIFU incompatibility. In view of these points, we used a single-channel coil in this study, but using the recently developed HIFU-compatible RF coil is expected to reduce this problem\cite{Correa2010_SciRep}. We can accommodate an acquired MR image obtained from multi-channel RF coils by simply increasing the number of network channels in the DeepACCnet. Moreover, it is also necessary to optimize the sampling scheme when considering the number of RF coils for signal reception. Furthermore, it seems possible to use a radial or spiral sampling strategy or echo planar imaging (EPI) for real-time MR temperature imaging. However, the additional problems that arise from these approaches, such as processing time and EPI ghosts, are challenges to be solved in the future.

It is known that VFA for measuring T1 is very sensitive to systematic errors from RF field inhomogeneities, slice profile, etc. \cite{Cheng2006_MRM}\cite{Braskute2018_ISMRM}. Moreover, accurate measurements of the parameters affected by T1 measurements, such as T1, T2*, M0, and FA, are challenging during heating treatment\cite{Graham1999_MRM}\cite{Kim2018_iMRI}. In this work, pre- and post-processing steps for T1 mapping were performed as described in the Materials and Methods section.  

One of the common and critical limitations of deep learning involves limited training datasets. Many previous studies have performed training data augmentation to prevent overfitting and increase accuracy\cite{Ciresan2010_NC}. In this study, 2D rotation, which is the simplest way to augment the available data, was used. Image-based methods, such as flipping, rotation, bending, and edge enhancement, have also performed well\cite{Taylor2017_Arxiv}. Recent studies have shown improved performance when performing data augmentation through physics-driven simulated data or synthetic generation\cite{Cho2019_MRM}\cite{Yoon2018_NI}. When water fat separation was performed by generating synthetic field inhomogeneity, incomplete water fat separation due to the field inhomogeneity could be overcome\cite{Cho2019_MRM}. Overall, it is expected that the proposed method will be able to perform data augmentation through simulated heat patterning. However, PRF-based MR thermometry and T1-based MR thermometry are phase-based and magnitude-based methods, respectively. Thus, in the proposed method, the variation due to synthetic heat can be simulated by generating complex MR images to perform data augmentation.

\section{Conclusion}

In this study, we demonstrated deep learning-based temperature imaging in both adipose and aqueous tissues, providing real-time interactive and high-quality imaging. This was validated based on agarose gel phantom, porcine muscle and porcine muscle with fatty layers heating tests, and in vivo prostate and brain non-heating tests. The proposed method could be critically useful in monitoring ablative therapies in both aqueous and adipose tissues.

\newpage     

\section{Tables}

\begin{table}[htbp]
\begin{center}
\captionv{10}{aaa}{Acquisition and processing times for fully sampled + curve fitting and cascaded CNN with high resolution.
\label{Table_1}
\vspace*{2ex}
}

\begin{tabular} {|p{3cm} p{3cm}|p{1.5cm}|p{1.5cm}|p{1.5cm}|p{1.5cm}|p{1.5cm}|}
\hline
\cline{4-7}
        & & Phantom & porcine Muscle & porcine Muscle with fat layers & In vivo human brain & In vivo human prostate   \\
\hline
        & Acquisition time & 5.6 s & 5.6 s & 5.6 s & 5.6 s & 5.6 s \\
Fully sampled + Curve Fitting &  &  &  &  & & \\
        & Processing time & 8 h & 8 h & 8 h & 8 h & 11 h   \\
\hline
        & Acquisition time & 0.8 s & 0.8 s & 0.8 s & 0.8 s & 0.8 s \\
Cascaded CNN with HR &  &  &  &  & & \\
        & Processing time & 32 ms & 32 ms & 32 ms & 32 ms & 78 ms \\
\hline

\end{tabular}
\end{center}
\end{table}


\clearpage
\section{Figures}

\clearpage
\section*{Acknowledgements}
This research was supported by the Technology Innovation Program (\#10076675) funded by the Ministry of Trade, Industry \& Energy (MOTIE, Korea). 

\section*{Conflicts of interest}
The authors have no relevant conflicts of interest to disclose.

\section*{References}
\addcontentsline{toc}{section}{\numberline{}References}
\vspace*{-20mm}


\bibliography{./MS19_MedPhys_RealTherm}      



\bibliographystyle{./medphy.bst}    


\clearpage
\section*{Supplementary material}
\newcommand{\beginsupplement}{%
        \setcounter{table}{0}
        \renewcommand{\thetable}{S\arabic{table}}%
        \setcounter{figure}{0}
        \renewcommand{\thefigure}{S\arabic{figure}}%
     }
     
\beginsupplement
\begin{figure}[ht]
   \begin{center}
   \includegraphics[width=14cm]{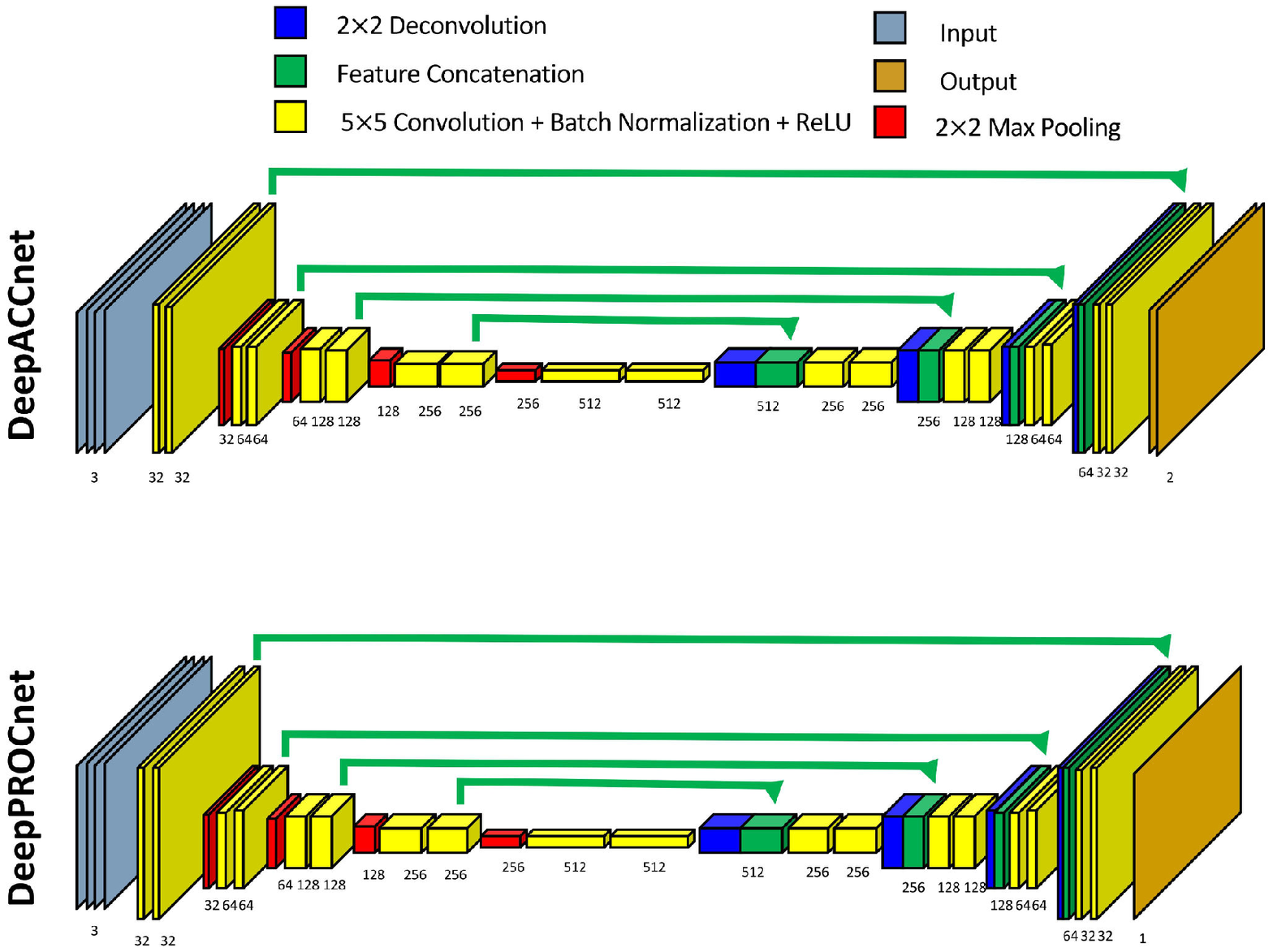}
   \captionv{12}{Short title - can be blank}{Simplified U-Net architectures for DeepACCnet and DeepPROCnet. The numbers of channels of the convolutional layers are summarized at the bottom of each block. ReLU: rectified linear unit.
   \label{SFigS1} 
    }  
    \end{center}
\end{figure}

\end{document}